\documentclass[pra,prl,twocolumn,showpacs,floatfix,10pt,twoside]{revtex4-1}
\usepackage{graphicx}
\usepackage{dcolumn}
\usepackage{bm}
\usepackage{subfigure}
\usepackage{float}
\usepackage{multirow}
\usepackage{booktabs}
\usepackage{amsmath}
\usepackage{latexsym,amssymb}
\usepackage[mathscr]{eucal}
\usepackage[switch*]{lineno}
\usepackage[bookmarks=true,
   colorlinks=true,
   linkcolor=blue,
   urlcolor=blue,
   citecolor=blue,
   bookmarks=true,
   hyperindex=true
]{hyperref}

\begin{document}


\title{Rainbow gravity corrections to the information f\/lux of a black hole and the sparsity of Hawking radiation}

\author{Zhong-Wen Feng\textsuperscript{1}}
\altaffiliation{Email: zwfengphy@163.com}
\author{Shu-Zheng Yang\textsuperscript{1}}
\altaffiliation{Email: szyangcwnu@126.com}
\vskip 0.5cm
\affiliation{1 Physics and Space Science College, China West Normal University, Nanchong, 637009, China}


\begin{abstract}
In this paper, we investigated the information f\/lux of Schwarzschild black hole and the sparsity of Hawking radiation in the presence of  rainbow gravity. The results demonstrate that the rainbow gravity has a very signif\/icant ef{f}ect on the information f\/lux. When the mass of rainbow black hole approaches to the order of Planck scale, the Bekenstein entropy loss per emit\/ted quanta in terms of the mass of black hole reduces to zero. Furthermore, we also f\/ind the sparsity of Hawking radiation in rainbow gravity is no longer a constant; instead, it monotonically decreases as the mass of black hole decrease. At the f\/inal stages of evaporation, contrary to those of GUP case,  the modif\/ied sparsity becomes inf\/inity, which indicates the ef\/fect of quantum gravity stops Hawking radiation and leads to remnant.
\end{abstract}
\maketitle
\section{Introduction}
\label{Int}
In the theoretical physics, one of the surprising achievements is to prove that black holes have thermodynamic properties \cite{cha1,cha2,cha3}. This discovery started out with an analogy connecting the laws of gravity and those of thermodynamics. In 1973, Bekenstein pointed out that the entropy of a black hole $S$ can be def\/ined in terms of horizon area $A$, namely, $S = {{Ak_B c^3 } \mathord{\left/ {\vphantom {{Ak_B c^3 } {4\hbar ^2 G}}} \right. \kern-\nulldelimiterspace} {4\hbar ^2 G}}$, with the Boltzmann constant $k_B$, the Planck constant $\hbar$, and the Newton's gravitational constant $G$ \cite{cha4}. Shortly afterwards, in refs.~\cite{cha5,cha6}, Hawking put forward the theory of black hole radiation, which is called as Hawking radiation now. The theory of Hawking radiation includes at least two novelties, one of which is showing the black holes radiate as black bodies, with characteristic temperature as $T = {\kappa  \mathord{\left/ {\vphantom {\kappa  {2\pi }}} \right. \kern-\nulldelimiterspace} {2\pi }}$, where $\kappa$ is the surface gravity of the black hole, and the other one indicates that the black hole is not the end of stellar evolution. Since Hawking radiation is radically af\/fected on the thermodynamic theory, the gravitational theory, and the quantum mechanism, it has received wide attention\cite{cha7,cha8,cha9,cha10}.

Despite most people have focused on using the Hawking radiation to analyze the thermodynamics of black holes in the past forty years, other properties of black holes can still be obtained by Hawking radiation such as the particle emission rates and information loss of black holes. As we know, the information of black holes can be ref\/lected by their three ``hairs'', namely, the mass $M$, charge $Q$, and angular momentum $\Omega$. As the black holes radiate the particles, their emission rates and information would change \cite{cha11}. Page f\/irst calculated the particle emission rates from an uncharged, non-rotating hole\cite{cha12}. Then, this work has been extended to other spacetimes \cite{cha13,cha14}. In refs.~\cite{cha15,cha16,cha17}, Alonso-Serrano and Visser quantif\/ied the information budget in evolution of black hole by considering that the entropy f\/lux of black holes is compensated by hidden information. Their results showed the lifetimes of black holes are related to the particle emission rates and information loss. Nevertheless, a lot of works claimed that the previous studies are not impeccable since the classical theory of Hawking radiation has some puzzles \cite{cha18,cha19,cha20,cha21}. For example, the black holes would evaporates completely into Hawking radiation since there is no cut-of\/fs, it makes the singularity of black holes exposed in the universe \cite{cha21}. If one further assumes the radiation is pure thermal, the black holes then lost all their information, which leads to ``Information loss paradox''. In order to solve those puzzles, the authors in refs.~\cite{cha20,cha21,cha22,cha23,cha24,cha25} analyzed how the black holes loss their information and how to recover it. In addition, the puzzles of black holes can be also solve by combining the models of quantum gravity with theory of Hawking radiation. According to the generalized uncertainty principle (GUP), which is a quantum gravity inspired correction to the Heisenberg's uncertainty principle at Planck scale, Adler et al. calculated the GUP corrections to the thermodynamic evolution of black holes \cite{cha26,cha27+,cha27,cha28,cha29,cha30}. Those results show the GUP can stop the evaporation of black holes and leads to remnant at the late stages of evolution, which indicates the GUP have an important ef\/fect on the information loss of black hoes. Therefore, the GUP corrections to the information f\/lux of black hole and the sparsity of Hawking radiation are investigated recently \cite{cha31,cha31+}. According to those modif\/ications, it is found that the information/entropy f\/lux is related to the mass of black hole, and the sparsity of Hawking radiation becomes thicker and thicker when a black hole approaches the Planck scale.

On the other hand, as the basis of loop quantum gravity (LQG), non-commutative geometry, spacetime discreteness, the standard energy-momentum dispersion relation would be changed to the so-called modif\/ied dispersion relation (MDR) when it approaches the Planck scale. Using the special relativity together with MDR, the double special relativity (DSR), which takes the speed of light $c$ and the Planck  scale as constants, has put forward by Amelino-Camelia \cite{cha32}. Subsequently, Magueijo and Smolin generalized the DSR to the curved spacetime, and arrive at the theory of rainbow gravity (or doubly general relativity) \cite{cha33}. In the theory of rainbow gravity (RG), the authors proposed the geometry of spacetime is related to the energy of the test particles. Hence, the background of this spacetime can be represented by a family of energy dependent metrics, namely, rainbow metrics. Now, the RG is considered as other promising candidates for a quantum gravity theory, which can modify the Hawking radiation and thermodynamic evolution of black holes just like GUP  \cite{cha34,cha34a1,cha34a2,cha34a3,cha34+,cha35+,cha36,cha37,cha38,cha39,cha40,cha40a1,cha40a2,cha40a3,cha40a5,cha40a6,cha41,cha42,cha44}. Besides those, one may f\/ind  that the  implications of the aspects of RG have been investigated in many contexts. For example, Hendi \emph{et al}. investigate the universe evolution without singularity by using the RG \cite{cha40a4,cha40a7}. In refs.~\cite{cha40b1,cha40b2,cha40b5}, the authors found the maximum mass and structure of neutron stars due to the RG corrected TOV equation. In ref.~\cite{cha40b3}, Gim and Gwak showed that the second law of thermodynamics and cosmic censorship conjecture are violated owing to the ef\/fect of RG \cite{cha40b2}. Channuie studed the deformed Starobinsky model in  the context of RG \cite{cha40b4}.

Due to the above discussion, it is f\/ind that both the GUP and RG have a very signif\/icant ef\/fect on radiations of black holes, which can be ref\/lected by information f\/lux. Therefore, inspired by the work in refs.~\cite{cha31,cha31+}, we calculate the RG corrections to the information f\/lux of Schwarzschild (SC) black hole and its sparsity in this paper. To begin with, incorporating the line element of SC black hole with the rainbow functions that were proposed by Amelino-Camelia et al., the rainbow SC black hole is constructed. Then, using the relation $E \ge {1 \mathord{\left/ {\vphantom {1 {r_H }}} \right. \kern-\nulldelimiterspace} {r_H }} = {1 \mathord{\left/ {\vphantom {1 {2GM}}} \right. \kern-\nulldelimiterspace} {2GM}}$, the RG corrected Hawking temperature entropy are obtained. Finally, according to these modif\/ication, the information f\/lux of rainbow SC black hole and the sparsity of Hawking radiation are analyzed.

The paper is organized as follows. In the next section, we brief\/ly review the thermodynamics of rainbow SC black hole. Section~\ref{SEI} is devoted to investigating the information f\/lux of rainbow SC black hole. In Section~\ref{SEII}, we discuss the RG corrected sparsity of Hawking radiation. Finally, the discussion and conclusion are presented in Section~\ref{Dis}.

\section{A brief on the thermodynamics of rainbow SC black hole}
\label{RF}
In this section, we brief\/ly review the thermodynamics properties of SC black hole in the RG. For obtaining these modif\/ications, it is necessary to constructs the rainbow
functions from the general form of MDR, which is
\begin{equation}
\label{eq1}
E^2 \mathcal{F}^2 \left( {{E \mathord{\left/ {\vphantom {E {E_p }}} \right. \kern-\nulldelimiterspace} {E_p }}} \right) - p^2 \mathcal{G}^2 \left( {{E \mathord{\left/ {\vphantom {E {E_p }}} \right. \kern-\nulldelimiterspace} {E_p }}} \right) = m^2 ,
\end{equation}
with the Planck energy $E_p$ and the energy of a text particle $E$,  and the correction terms $\mathcal{F}\left( {{E \mathord{\left/{\vphantom {E {E_p }}} \right. \kern-\nulldelimiterspace} {E_p }}} \right) $ and $\mathcal{G}\left( {{E \mathord{\left/ {\vphantom {E {E_p }}} \right. \kern-\nulldelimiterspace} {E_p }}} \right)$ are known as rainbow functions, which are required to satisfy the relationship $\mathop {\lim }\limits_{{E \mathord{\left/ {\vphantom {E {E_p }}} \right. \kern-\nulldelimiterspace} {E_p }} \to 0} \mathcal{F}\left( {{E \mathord{\left/ {\vphantom {E {E_p }}} \right. \kern-\nulldelimiterspace} {E_p }}} \right) = 1$ and $\mathop {\lim }\limits_{{E \mathord{\left/ {\vphantom {E {E_p }}} \right. \kern-\nulldelimiterspace} {E_p }} \to 0} \mathcal{G}\left( {{E \mathord{\left/  {\vphantom {E {E_p }}} \right. \kern-\nulldelimiterspace} {E_p }}} \right) = 1$. In this case, eq.~(\ref{eq1}) goes to the standard energy-momentum dispersion relation  at low energy scale, that is,  $E^2  - p^2  = m^2$.

In the literature of RG, the forms of rainbow functions are based on dif\/ferent phenomenological motivations. Many forms of rainbow functions are referred to in
refs.~\cite{chb0,chb0+,chb2+} and references therein. In this work, we employ one of the most interesting rainbow functions that was
proposed by Amelino-Camelia et al. \cite{chb1,chb2}. Among the $\kappa$-Minkowski non-commutative geometry and LQG, Amelino-Camelia et al.
constructed a form of MDR in the high-energy regime, which takes the form as
\begin{equation}
\label{eq1+}
E^2  - \vec p^2  + \eta \vec p^2 \left( {{E \mathord{\left/ {\vphantom {E {E_p }}} \right. \kern-\nulldelimiterspace} {E_p }}} \right)^4  \simeq m^2.
\end{equation}
where the third term in the RHS is the correction term. When $\eta=0$ or $ {{E \mathord{\left/ {\vphantom {E {E_p }}} \right. \kern-\nulldelimiterspace} {E_p }}} \rightarrow0$, the MDR becomes the standard energy-momentum dispersion relation $E^2  - \vec p^2   = m^2$. Meanwhile, since the energy of a particle can hardly exceed the Planck energy, we have $0 < {E \mathord{\left/ {\vphantom {E {{E_p}}}} \right. \kern-\nulldelimiterspace} {{E_p}}} \leq 1$.
Now, comparing this MDR~(\ref{eq1+}) with eq.~(\ref{eq1}), the rainbow functions can be expressed in the following form:
\begin{equation}
\label{eq2}
\mathcal{F}\left( {{E \mathord{\left/ {\vphantom {E {E_p }}} \right. \kern-\nulldelimiterspace} {E_p }}} \right) = 1, \quad
\mathcal{G}\left( {{E \mathord{\left/ {\vphantom {E {E_p }}} \right. \kern-\nulldelimiterspace} {E_p }}} \right) = \sqrt {1 - \eta \left( {\frac{E}{{E_p }}} \right)^4 } ,
\end{equation}
where $\eta$ is the rainbow parameter. Despite the absence of lower bound for the rainbow parameter $\eta$, one still can  analyze the upper bound of $\eta$  by using various experimental considerations \cite{chb2+,chb2a+}. Theoretically, it is always assumed  that $\eta$  is  of  the  order  of  unity, in which case the corrections are negligible unless energies approach the Planck energy $E_p$. However, when $\eta=0$, quantum gravity ef\/fect would disappear. Hence, in this work, we consider three cases, that is, $\eta>0$, $\eta=0$ and $\eta<0$.

In refs.~\cite{cha33,chb3}, the authors showed that the modif\/ied metric in gravity's rainbow can be obtained by replacing $dt \to {{dt} \mathord{\left/ {\vphantom {{dt} {\mathcal{F}\left( {{E \mathord{\left/ {\vphantom {E {E_p }}} \right. \kern-\nulldelimiterspace} {E_p }}} \right)}}} \right. \kern-\nulldelimiterspace} {\mathcal{F}\left( {{E \mathord{\left/ {\vphantom {E {E_p }}} \right. \kern-\nulldelimiterspace} {E_p }}} \right)}}$  for time coordinates and $dx^i  \to {{dx^i } \mathord{\left/ {\vphantom {{dx^i } {\mathcal{G}\left( {{E \mathord{\left/ {\vphantom {E {E_p }}} \right. \kern-\nulldelimiterspace} {E_p }}} \right)}}} \right. \kern-\nulldelimiterspace} {\mathcal{G}\left( {{E \mathord{\left/ {\vphantom {E {E_p }}} \right. \kern-\nulldelimiterspace} {E_p }}} \right)}}$  for all spatial coordinates. Hence, the metric in f\/lat spacetime is given by $ds^2  =  - {{dt^2 } \mathord{\left/ {\vphantom {{dt^2 } {{\cal F}\left( {{E \mathord{\left/ {\vphantom {E {E_p }}} \right. \kern-\nulldelimiterspace} {E_p }}} \right)}}} \right. \kern-\nulldelimiterspace} {{\cal F}\left( {{E \mathord{\left/ {\vphantom {E {E_p }}} \right. \kern-\nulldelimiterspace} {E_p }}} \right)}}^2  + {{dx_i {\rm{d}}x^i } \mathord{\left/ {\vphantom {{dx_i {\rm{d}}x^i } {{\cal G}\left( {{E \mathord{\left/ {\vphantom {E {E_p }}} \right. \kern-\nulldelimiterspace} {E_p }}} \right)}}} \right. \kern-\nulldelimiterspace} {{\cal G}\left( {{E \mathord{\left/ {\vphantom {E {E_p }}} \right.
 \kern-\nulldelimiterspace} {E_p }}} \right)}}^2$, and the most studied spherically symmetric metric takes the form as follows \cite{cha33}:
\begin{align}
\label{eq3}
ds^2  = & - \frac{{A\left( r \right)}}{{{\cal F}\left( {{E \mathord{\left/ {\vphantom {E {E_P }}} \right. \kern-\nulldelimiterspace} {E_P }}} \right)^2 }}dt^2  + \frac{{B\left( r \right)^{ - 1} }}{{{\cal G}\left( {{E \mathord{\left/ {\vphantom {E {E_P }}} \right. \kern-\nulldelimiterspace} {E_P }}} \right)^2 }}dr^2
 \nonumber \\
& + \frac{{r^2 }}{{{\cal G}\left( {{E \mathord{\left/ {\vphantom {E {E_P }}} \right. \kern-\nulldelimiterspace} {E_P }}} \right)^2 }}d\Omega ^2 ,
\end{align}
where $A\left( r \right) = B\left( r \right) = 1 - {{2Gk_B M} \mathord{\left/ {\vphantom {{2Gk_B M} {c^3 \hbar r}}} \right. \kern-\nulldelimiterspace} {c^3 \hbar r}}$
and $d\Omega ^2$ is the metric of two-dimensional unit sphere, respectively. By using the null hypersurface condition $g^{\mu \nu } \left( {{{\partial F} \mathord{\left/ {\vphantom
{{\partial F} {\partial x^\mu  }}} \right. \kern-\nulldelimiterspace} {\partial x^\mu  }}} \right)\left( {{{\partial F} \mathord{\left/ {\vphantom {{\partial F} {\partial x^\nu  }}} \right.
\kern-\nulldelimiterspace} {\partial x^\nu  }}} \right) = 0$, one can easily obtain the event horizon of rainbow SC black hole is $r_H  = {{2Gk_B M} \mathord{\left/ {\vphantom {{2Gk_B M}{c^3 \hbar }}} \right. \kern-\nulldelimiterspace} {c^3 \hbar }}$.
For a spherically symmetric spacetime, the original Hawking temperature satisf\/ies the following expression
\begin{align}
\label{eq4}
T_H &  = \frac{{\kappa _H }}{{2\pi }} = \left. {\frac{1}{{2\pi }}\sqrt { - \frac{1}{2}\nabla ^\mu  \xi ^\nu  \nabla _\mu  \xi _\nu  } } \right|_{r = r_H }
\nonumber \\
& =\frac{{c^3 \hbar }}{{8\pi GMk_B }},
\end{align}
where ${\kappa _H } = {1 \mathord{\left/ {\vphantom {1 {4GM}}} \right. \kern-\nulldelimiterspace} {4GM}}$ and ${\xi _\nu  }$ represent the surface gravity on the event horizon of
rainbow SC black hole and the time-like Killing vectors, respectively \cite{chb4,chb4+,chb5+,chb6+,chb7+}. Next, according to the rainbow functions eq.~(\ref{eq2}), the Hawking temperature of rainbow
SC black hole is given by
\begin{align}
\label{eq5}
T_H^{{\rm{RG}}}  &= \frac{\kappa_H^{{\rm{RG}}} }{{2\pi }} = \frac{{{\cal G}\left( {{E \mathord{\left/ {\vphantom {E {E_P }}} \right. \kern-\nulldelimiterspace} {E_P }}} \right)}}{{{\cal F}\left( {{E \mathord{\left/ {\vphantom {E {E_P }}} \right. \kern-\nulldelimiterspace} {E_P }}} \right)}}\frac{{\sqrt {\partial _r A\left( {r_H } \right)\partial _r B\left( {r_H } \right)} }}{{4\pi }}
\nonumber \\
& = \frac{{c^3 \hbar }}{{8\pi GMk_B }}\sqrt {1 - \eta \left( {\frac{E}{{E_p }}} \right)^4 },
\end{align}
where the  ${\kappa_H^{{\rm{RG}}} }  =  - \frac{1}{2}\mathop {\lim }\limits_{r \to r_H } \sqrt { - \frac{{g^{11} }}{{g^{00} }}} \frac{{\left( {g^{00} } \right)^\prime  }}{{g^{00} }}  = \frac{{\kappa _H \mathcal{G}\left( {{E \mathord{\left/ {\vphantom {E {E_p }}} \right. \kern-\nulldelimiterspace} {E_p }}} \right)}}{{2\pi \mathcal{F}\left( {{E \mathord{\left/  {\vphantom {E {E_p }}} \right. \kern-\nulldelimiterspace} {E_p }}} \right)}}$ is the modif\/ied surface gravity.
Following the argument in refs.~\cite{cha26,cha34,chb5}, the Heisenberg uncertainty principle $\Delta x$ $\Delta p \ge \hbar c$ can be translated to
a lower bound on the energy of radiant particle $E \ge {\hbar c\mathord{\left/ {\vphantom {\hbar c{\Delta x}}} \right. \kern-\nulldelimiterspace} {\Delta x}}$ with the uncertainty
position $\Delta x$. When considering the minimum value of $\Delta x$ equals to the event horizon $r_H$, one has
\begin{equation}
\label{eq6}
E \ge {{\hbar c} \mathord{\left/ {\vphantom {{\hbar c} {\Delta x}}} \right. \kern-\nulldelimiterspace} {\Delta x}} \approx {{\hbar c} \mathord{\left/ {\vphantom {{\hbar c} {2r_H }}} \right. \kern-\nulldelimiterspace} {2r_H }} = {{\hbar ^2 c^4 } \mathord{\left/ {\vphantom {{\hbar ^2 c^4 } {4GMk_B }}} \right.\kern-\nulldelimiterspace} {4GMk_B }}.
\end{equation}
Now, substituting eq.~(\ref{eq6}) into eq.~(\ref{eq5}), the rainbow temperature can be rewritten as \cite{cha34,chb6}
\begin{equation}
\label{eq7}
T_H^{\rm{RG}}  = \frac{{c^3 \hbar }}{{8\pi GMk_B }}\sqrt {1 - \eta \left( {\frac{{\hbar ^2 c^4 }}{{4GMk_B E_p }}} \right)^4 }.
\end{equation}
\begin{figure}
\centering 
\includegraphics[width=.52\textwidth,origin=c,angle=0]{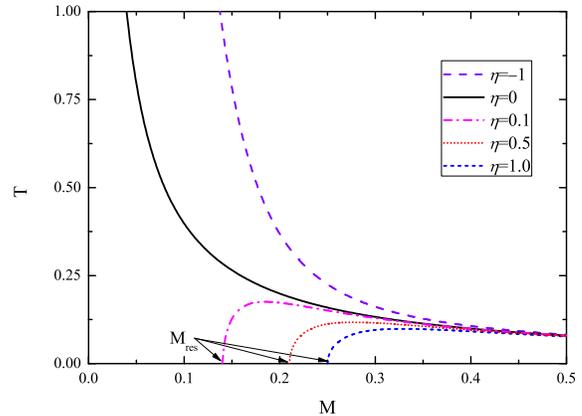}
\caption{\label{fig0} Hawking temperature of the mass of SC black hole for dif\/ferent $\eta$. Here, we take $G = \hbar  = k_B  = E_p  = 1$,}
\end{figure}
Mathematically, the modif\/ication are not only related to the original thermodynamic quantities, but also to the Planck length $E_p$,  the rainbow parameter  $\eta$. As it is obvious from f\/igure~\ref{fig0}, when $\eta= 0$, the modif\/ied temperature (black solid curve) reduces to the original case. For $\eta> 0$ (the blue dashed curve, red dotted curve and pink dot-dashed curve), it is clear that the ef\/fect of RG prevents the complete evolution of the black hole and leads to the remnant when the mass of the rainbow SC black hole decreases to a value of Planck scale, that is to say, the end point of the evolution of black holes is no longer zero, but a Planck scale. By solving eq.~(\ref{eq7}), the remnant mass is ${M_{{\rm{res}}}} = {{{c^4}{\hbar ^2}{\eta ^{1/4}}} \mathord{\left/ {\vphantom {{{c^4}{\hbar ^2}{\eta ^{1/4}}} {4G{k_B}{E_p}}}} \right. \kern-\nulldelimiterspace} {4G{k_B}{E_p}}}$, which can be considered as a candidate of dark matter or the ``Planck star'' \cite{cha34+,cha35+}. However, if the RG parameter takes a negative value, e.g. $\eta=-1$ for purple dashed curve, the second term of the RHS in eq.~(\ref{eq7}) is always be greater than zero. It implies that the ef\/fect of RG can accelerate the evaporation of black hole, which still leads to the information paradox of black holes since the increase of Hawking temperature is not limited. Despite the heated debate over the remain of black holes, but obtaining an object with limited mass that can store information to end the evaporation of black holes is  more palatable than having a divergent temperature as in the usual picture of Hawking evaporation \cite{cha20,chc2}. Therefore, for solving the catastrophic behavior of Hawking temperature, and study how the ef\/fect  of quantum gravity af\/fects the information f\/lux of black hole and and the sparsity of Hawking radiation, one should choice $\eta>0$ in the following research.

\section{The information f\/lux of rainbow SC black hole}
\label{SEI}
\begin{figure}
\centering 
\includegraphics[width=.52\textwidth,origin=c,angle=0]{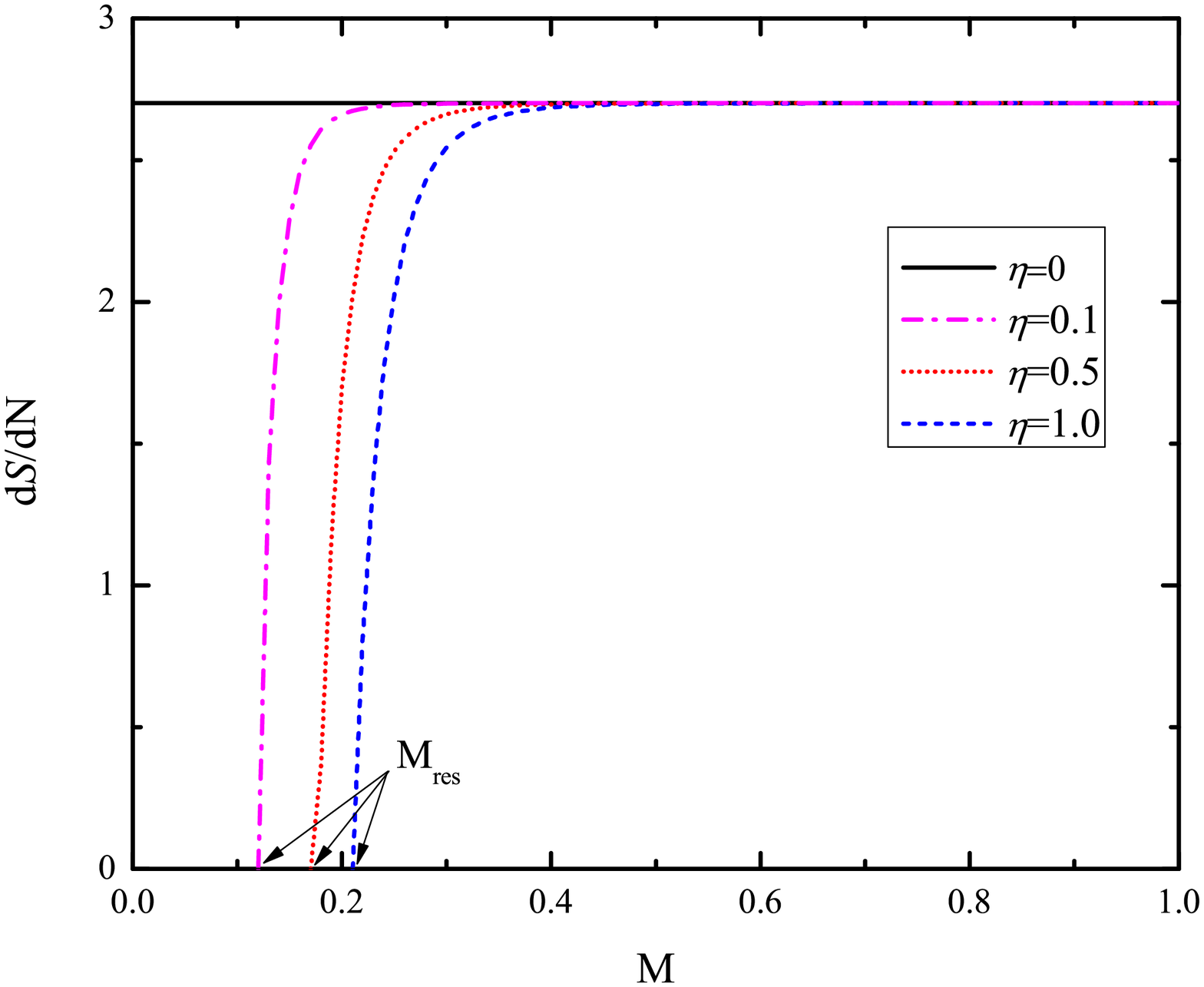}
\caption{\label{fig1} Bekenstein entropy loss per emit\/ted quanta in terms of the mass of SC black hole for dif\/ferent $\eta$. Here, we choose natural units $G = \hbar  = k_B  = E_p  = 1$.}
\end{figure}

According to the viewpoint in ref.~\cite{chc1}, it is feasible to assume an exact Planck spectrum at the Hawking temperature and the thermodynamic entropy is related to
the lack of information. Therefore, in order to analyze the information f\/lux of rainbow SC black hole, it is necessary to calculate the Bekenstein entropy loss per
emit\/ted quanta in terms of the mass and total number of emit\/ted particles. Based on the f\/irst law of black hole thermodynamics $c^2 dM = TdS$ and eq.~(\ref{eq7}), the entropy of rainbow SC black hole becomes
\begin{align}
\label{eq10}
{S^{RG}}&={S_0}\sqrt {1 - \eta {{\left( {\frac{{\pi {c^7}{\hbar ^3}}}{{4G{k_B}E_p^2{S_0}}}} \right)}^2}}
\nonumber \\
&= {S_0} - \frac{\eta }{{2{S_0}}}{\left( {\frac{{{c^7}{\hbar^3}\pi }}{{4G{k_B}E_p^2}}} \right)^2} +  {\cal O}\left( \eta  \right),
\end{align}
with the  the entropy of original SC black hole $S_0  = {{4\pi GM^2 k_B } \mathord{\left/ {\vphantom {{4\pi GM^2 k_B } {\hbar c}}} \right. \kern-\nulldelimiterspace} {\hbar c}}$. With the help of  eq.~(\ref{eq10}), the modif\/ied Bekenstein entropy loss of rainbow is given by
SC black hole per emit\/ted quanta \cite{cha17,chc2}
\begin{equation}
\label{eq11}
\frac{{d{S^{{\rm{RG}}}}}}{{dN}} = \frac{{{{d{S_0}} \mathord{\left/ {\vphantom {{d{S_0}} {dt}}} \right. \kern-\nulldelimiterspace} {dt}}{\rm{ }}}}{{{{dN} \mathord{\left/
 {\vphantom {{dN} {dt}}} \right.  \kern-\nulldelimiterspace} {dt}}}}\left[ {1 + \frac{\eta }{{2S_0^2}}{{\left( {\frac{{{c^7}{\hbar ^3}\pi }}{{4G{k_B}E_p^2}}} \right)}^2} + {\cal O}\left( \eta  \right)} \right],
\end{equation}
where $N$ is the number of particles, and Bekenstein entropy loss of original SC black hole per emit\/ted quanta is
\begin{equation}
\label{eq12}
\frac{{dS_0}}{{dN}} = \frac{{{{dS_0 } \mathord{\left/ {\vphantom {{dS_0 } {dt}}} \right. \kern-\nulldelimiterspace} {dt}}}}{{{{dN} \mathord{\left/
 {\vphantom {{dN} {dt}}} \right. \kern-\nulldelimiterspace} {dt}}}} = \frac{{8\pi k_B M}}{{c^2 m_p^2 }}\hbar \left\langle \omega  \right\rangle.
\end{equation}
With the help of the def\/inition of Bekenstein entropy and the conservation of energy  $\left\langle E \right\rangle  = \hbar \left\langle \omega  \right\rangle  =
{{\pi ^4 k_B T_H^{\rm{RG}} } \mathord{\left/ {\vphantom {{\pi ^4 k_B T_H^{\rm{RG}} } {30\zeta \left( 3 \right)}}} \right. \kern-\nulldelimiterspace} {30\zeta \left( 3 \right)}}$,
eq.~(\ref{eq12}) can be rewritten as
\begin{equation}
\label{eq13}
\frac{{dS_0 }}{{dN}} = \frac{{k_B \pi ^4 }}{{30\zeta \left( 3 \right)}}\frac{{8\pi k_B M}}{{c^2 m_p^2 }}T_H^{\rm{RG}}.
\end{equation}
Substituting the RG corrected Hawking temperature  into eq.~(\ref{eq13}) and expanding, the Bekenstein entropy loss of original SC black hole per emit\/ted quanta is given by
\begin{equation}
\label{eq14}
\frac{{d{S_0}}}{{dN}} = \frac{{{k_B}{\pi ^4}}}{{30\zeta \left( 3 \right)}}\left[ {1 - \frac{\eta }{2}{{\left( {\frac{{{c^4}{\hbar^2}}}{{4G{k_B}{E_p}M}}} \right)}^4} + {\cal O}\left( \eta  \right)} \right].
\end{equation}
Then, substituting eq.~(\ref{eq14}) into eq.~(\ref{eq11}), the Bekenstein entropy loss of rainbow SC black hole per emit\/ted quanta becomes
\begin{equation}
\label{eq15}
\frac{{d{S^{{\rm{RG}}}}}}{{dN}} = \frac{{{k_B}{\pi ^4}}}{{30\zeta \left( 3 \right)}}\left[ {1 - {{\left( {\frac{\eta }{2}} \right)}^2}{{\left( {\frac{{{c^4}{\hbar ^2}}}{{4G{k_B}M{E_p}}}} \right)}^8} +{\cal O} \left( \eta  \right)} \right].
\end{equation}
From abovementioned equation, one can plot the Bekenstein entropy loss per emit\/ted quanta in terms of the mass of SC black hole for dif\/ferent values of rainbow parameter $\eta$ in f\/igure~\ref{fig1}.

In f\/igure~\ref{fig1},  if $\eta=0$,  one can observe that the Bekenstein entropy loss of original SC black hole per emit\/ted quanta ${{dS_0 } \mathord{\left/ {\vphantom {{dS_0 } {dN}}} \right. \kern-\nulldelimiterspace} {dN}}$(black solid line) is a constant, which is about $2.70$ \cite{cha15,cha16,cha17}. The blue dashed line, red dotted line and pink dot-dashed line in the diagram illustrate the Bekenstein entropy loss of rainbow SC black hole per emit\/ted quanta ${{dS^{\rm{RG}}} \mathord{\left/ {\vphantom {{dS} {dN}}} \right. \kern-\nulldelimiterspace} {dN}}$, the value of rainbow parameter  $\eta$ decreases from bottom to top. When mass of the black hole is large enough, the behavior of  ${{dS^{\rm{RG}}} \mathord{\left/ {\vphantom {{dS} {dN}}} \right. \kern-\nulldelimiterspace} {dN}}$ is similar to that of the original case, it implies the ef\/fect of RG is negligible at big scale. However, the behavior of Bekenstein entropy loss of rainbow SC black hole per emit\/ted quanta is apart from that of the original case with the development of evolution. It is clear that ${{dS^{\rm{RG}} }\mathord{\left/ {\vphantom {{dS} {dN}}} \right. \kern-\nulldelimiterspace} {dN}}$ monotonically decreases in mass. Furthermore, when the mass of rainbow SC black hole approaches the Planck mass, the ${{dS^{\rm{RG}}} \mathord{\left/ {\vphantom {{dS} {dN}}} \right. \kern-\nulldelimiterspace} {dN}}$ reaches zero. By solving  eq.~(\ref{eq15}) and considering the approximation we made in eq.~(\ref{eq10}), one has $M_{{\rm{min}}} \simeq {M_{{\rm{res}}}} $, where $M_{{\rm{min}}}$ is the value of zero mass point in f\/igure~\ref{fig1}.  Therefore, by analyzing the information f\/lux of SC black hole, one can still f\/ind the same results as those in ref.~\cite{cha34,cha34a1,cha34a2,cha34a3},  that is, the ef\/fect of RG stops Hawking radiation in the f\/inal stages of black holes' evolution and leads to the remnant. With the disappearance of Hawking radiation, the rainbow SC black hole no longer exchange information with its surroundings, which causes the information of rainbow SC black holes is enclosed in the remnant. Meanwhile, for  $M < {M_{{\rm{min}}}}$, all the thermodynamic quantities violate the f\/irst law of thermodynamics. Therefore,  ${{dS^{\rm{RG}} }\mathord{\left/ {\vphantom {{dS} {dN}}} \right. \kern-\nulldelimiterspace} {dN}}$  does not make sense in that region.

\begin{figure}
\centering 
\includegraphics[width=.52\textwidth,origin=c,angle=0]{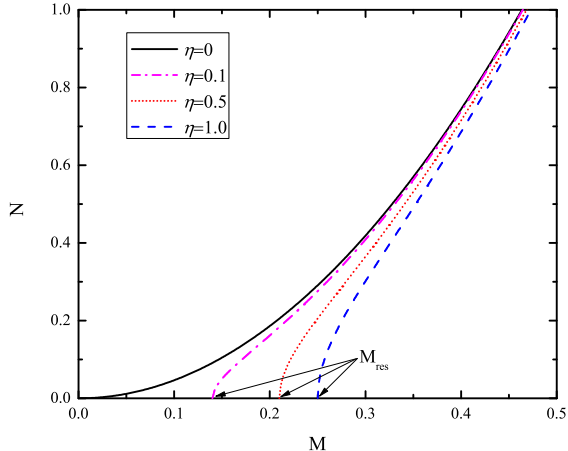}
\caption{\label{fig2} Total number of particles emit\/ted from SC black hole as a function of $M$ for dif\/ferent values $\eta$. We choose natural units $G = \hbar  = k_B  = E_p  = 1$.}
\end{figure}

Next, the modif\/ied total number of emit\/ted quanta can be expressed as follows:
\begin{equation}
\label{eq16}
\frac{{dN}}{{dM}} = \frac{{30\zeta \left( 3 \right)c^2 }}{{k_B \pi ^4 T_H^{\rm{RG}} }}.
\end{equation}
According to eq.~(\ref{eq7}), the total number of particles emit\/ted from rainbow SC black hole is given as follows:
\begin{equation}
\label{eq17}
N = \frac{{30\zeta \left( 3 \right)}}{{{\pi ^4}}}\left[ {\frac{{4\pi G{M^2}}}{{c\hbar }} - \eta \frac{{\pi {c^{15}}{\hbar ^7}}}{{128{G^3}k_B^4{M^2}E_p^4}}}+ {\cal O} \left( \eta  \right) \right].
\end{equation}
where $M$ is the initial mass of rainbow SC black hole. In ref.~\cite{cha31}, the entropy in nats is $\hat S = {{S_0 } \mathord{\left/ {\vphantom {{S_0 } {k_B }}} \right.
\kern-\nulldelimiterspace} {k_B }} = {{4\pi GM^2 } \mathord{\left/ {\vphantom {{4\pi GM^2 } {c\hbar }}} \right. \kern-\nulldelimiterspace} {c\hbar }}$.
In this case, eq.~(\ref{eq17}) can expressed in terms of the entropy in nats, that is
\begin{equation}
\label{eq18}
N = \frac{{30\zeta \left( 3 \right)}}{{{\pi ^4}}}\left[ {\hat S - \frac{\eta }{{32}}{{\left( {\frac{{{c^7}{\hbar ^3}\pi }}{{Gk_B^2E_p^2}}} \right)}^2}\frac{1}{{\hat S}} + {\cal O} \left( \eta  \right)} \right],
\end{equation}
 It should be noted that the original total number of particles emit\/ted from a black hole $N_0  = {{30\zeta \left( 3 \right)\hat S} \mathord{\left/  {\vphantom {{30\zeta \left( 3 \right)\hat S} {\pi ^4 }}} \right.\kern-\nulldelimiterspace} {\pi ^4 }}$. Obviously, eq.~(\ref{eq18})  shows that the quantum gravity ef\/fect can reduce total number of particles emit\/ted from black hole. Moreover, by setting $G=\hbar  = k_B  = E_p  = 1$, the total number of particles emit\/ted from SC black hole as a function of $\eta$ is plot in f\/igure~\ref{fig2}.

As one can see from f\/igure~\ref{fig2}, the blue dashed line, red dotted line and purple dot-dashed line illustrate the RG corrected total number of particles emit\/ted $N$, whereas original cases is represented by the black solid line. At the early stage of black hole evolution, these lines coincide together. After that, the total numbers of particles emit\/ted are gradually reduced via the Hawking radiation, the original total number of particles emit\/ted $N_0$ vanishes when $M \to 0$, while the RG corrected one reduce to zero mass point at $M_{{\rm{min}}}$. Utilizing eq.~(\ref{eq17}), one can f\/ind the relation $M_{{\rm{min}}} \simeq M_{{\rm{res}}} $, which implies that the quantum gravity ef\/fect can obviously af\/fect the evolution of black holes.

\section{The RG corrected sparsity of Hawking radiation}
\label{SEII}
\begin{figure}
\centering 
\includegraphics[width=.52\textwidth,origin=c,angle=0]{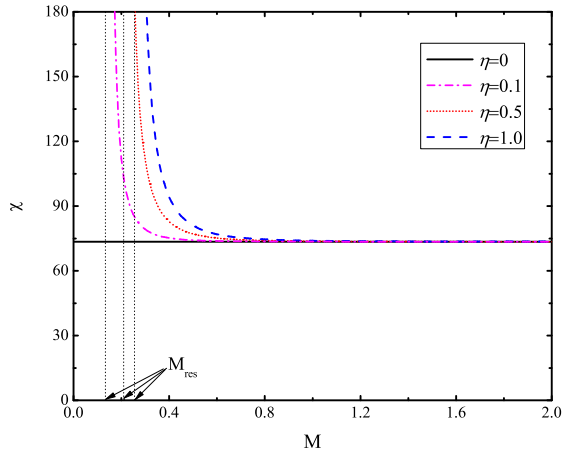}
\caption{\label{fig3} The sparsity of Hawking radiation as a function of $M$ for dif\/ferent values $\eta$. We choose natural units $G = \hbar  = k_B  = E_p  = 1$.}
\end{figure}

Another important property of Hawking radiation is its sparsity. In ref.~\cite{chc1}, the authors showed that the sparsity of Hawking radiation can be describe by
a dimensionless parameter  $\chi$, which is the ratio between an average time between the emission of two consecutive quanta and the natural time scale.
The result implies that the sparsity of Hawking radiation is thin during the whole evaporation process. Recently, researches show that Hawking radiation is no longer sparse via
the ef\/fect of GUP with positive parameter \cite{cha31}. However, when considering the negative GUP parameter, the sparsity of Hawking radiation would be enhanced \cite{chc2}.
Therefore, it is interesting to analyze how does RG af\/fect the sparsity of Hawking radiation. Now, the expression of dimensionless parameters is given by
\begin{equation}
\label{eq19}
\chi  = \frac{{C\lambda _{{\rm{thermal}}}^2}}{{g{A_{{\rm{effective}}}}}},
\end{equation}
where  $C$ represent a dimensionless constant that depends on the specif\/ic parameter $\chi$ we are choosing,  $g$ is the spin degeneracy factor,
$A$ is the ef\/fective area and $\lambda _{{\rm{thermal}}}  = {{2\pi \hbar c} \mathord{\left/ {\vphantom {{2\pi \hbar c} {k_B T_0 }}} \right. \kern-\nulldelimiterspace} {k_B T_0 }}$
is the thermal wavelength of a Hawking particle, respectively. In ref.~\cite{chc2}, the author pointed out that $\chi  \ll 1$ denotes a typical blackbody radiation, which implies that the black holes emit particles continuously. On the other hand, the Hawking radiation becomes extremely sparse for $\chi  \gg 1$. It is well known that the area and horizon radius of SC black hole satisfying relationship ${1 \mathord{\left/ {\vphantom {1 {4A_H }}} \right. \kern-\nulldelimiterspace} {4A_H }} = \pi r_H^2$. However, this relationship is actually only applicable to some certain types of
particles in the low frequency limit. For the high frequencies cases, the relationship between area and horizon radius of SC black hole becomes ${{27A_H } \mathord
{\left/ {\vphantom {{27A_H } {16}}} \right. \kern-\nulldelimiterspace} {16}} = {{27\pi r_H^2 } \mathord{\left/ {\vphantom {{27\pi r_H^2 } 4}} \right. \kern-\nulldelimiterspace} 4}$
with the enhancement factor ${{27} \mathord{\left/ {\vphantom {{27} 4}} \right.\kern-\nulldelimiterspace} 4}$ \cite{chc1}. Hence, the ef\/fective area of original
SC black hole in the high frequency limit is def\/ined by $A_{{\rm{ef\/fective}}}  = {{27A_H } \mathord{\left/ {\vphantom {{27A_H } 4}} \right. \kern-\nulldelimiterspace} 4}$.
Substituting the original thermal wavelength and ef\/fective area into eq.~(\ref{eq19}), the original relevant factor in any dimensionless parameter for massless bosons is
\begin{equation}
\label{eq20}
\chi_0  = \frac{{\lambda _{{\rm{thermal}}}^2 }}{{A_{{\rm{ef\/fective}}} }} = \frac{{64\pi ^3 }}{{27}} \approx 73.5.
\end{equation}
 It is clear that eq.~(\ref{eq20}) is related to the properties of a black hole, which leads to the f\/inal result being a constant,
 it means the sparsity never change during the whole evaporation process. However, when considering the ef\/fect of RG,
 the modif\/ied ef\/fective area can be expressed as following:
\begin{equation}
\label{eq21}
A_{{\rm{effective}}}^{{\rm{RG}}} = \frac{{27{A_H}}}{4}\sqrt {1 - \eta \frac{{{\pi ^2}{c^8}{\hbar ^8}}}{{A_H^2k_B^4E_p^4}}} ,
\end{equation}
and the modif\/ied thermal wavelength becomes
\begin{align}
\label{eq22}
\lambda _{{\rm{thermal}}}^{{\rm{RG}}}  & = \frac{{2\pi \hbar c}}{{k_B T_H^{\rm RG} }}
\nonumber \\
& = \frac{{2\pi c\hbar }}{{k{T_H}}}{\left[ {1 - \eta {{\left( {\frac{{2\pi c\hbar }}{{{E_p}}}{T_H}} \right)}^4}} \right]^{ - \frac{1}{2}}} .
\end{align}
It should be noted that we did not expand the temperature and entropy expressions in the above derivations to ensure the accuracy of the f\/inal results. According eq.~(\ref{eq21}) and eq.~(\ref{eq22}), the RG corrected dimensionless parameter is given by
\begin{align}
\label{eq23}
{\chi ^{{\rm{RG}}}} &= \frac{{\left( {\lambda _{{\rm{thermal}}}^{{\rm{RG}}} } \right)^2 }}{{A_{{\rm{ef\/fective}}}^{{\rm{RG}}} }}
 \nonumber \\
 &=\frac{{64{\pi ^3}}}{{27}}{\left[ {1 - \eta {{\left( {\frac{{{c^4}{\hbar ^2}}}{{4Gk_B^{}ME_p^{}}}} \right)}^4}} \right]^{ - \frac{3}{2}}}.
\end{align}
Dif\/ferent from the original case $\chi_0$, eq.~(\ref{eq23}) shows that the modif\/ied dimensionless parameters are not only dependent on the ratio ${{64\pi ^3 } \mathord{\left/  {\vphantom {{64\pi ^3 } {27}}} \right. \kern-\nulldelimiterspace} {27}}$, but also determined by the mass of black hole $M$, Planck energy $E_p$ and RG parameter $\eta$. Moreover, when $M \rightarrow M_{{\rm{res}}}$, the denominator of eq.~(\ref{eq23})  tends to zero,  so the ${\chi ^{{\rm{RG}}}}$ are diverges at $M_{{\rm{res}}}$. In order to discuss the behaviors of dimensionless parameter $\chi$, we plot f\/igure~\ref{fig3}.

As seen from f\/igure~\ref{fig3},  one can f\/ind that the blue dashed curve, red dotted curve and pink dot-dashed curve for modif\/ied dimensionless parameter $\chi ^{{\rm{RG}}}$ diverges when the mass approaches $M_{{\rm{res}}}$ (the verticals solid lines), which is dif\/ferent from the black solid line for original dimensionless parameter $\chi_0$ that keeps a constant value during the whole evaporation process. From the above discussion, it can be concluded that the sparsity of Hawking radiation is enhanced due to the ef\/fect of RG; hence, the pause time between in Hawking radiation becomes longer and longer. The rainbow SC black hole takes inf\/inite time to radiate a particle when the black hole at the f\/inal stages of evaporation. In other words, the rainbow SC black hole does not radiate any particle or lose its information when $M \rightarrow M_{{\rm{res}}}$.

 It is interesting to compare our results with those of GUP cases. In ref.~\cite{chc2}, one f\/inds that the modif\/ied radiation becomes inf\/initely sparse with negative GUP parameter, which is  remains qualitatively the same as our work. Therefore, we think the there is a deeper connection between RG and GUP model with negative parameter. In particular, Ong recently claimed that the negative GUP parameter can ensure the validity of Chandrasekhar limit \cite{chc3}. According to refs.~\cite{chc4,chc5,chc6} and the our discussion in this work, it is believe that the RG can also ensure the validity of Chandrasekhar limit and af\/fect the white dwarf physics. But when considering the GUP model with positive parameter \cite{cha31+}, the GUP corrected dimensionless parameters is dif\/ferent from our case. Hence, we think the dif\/ference originate from the thermal radiation behavior of black hole under dif\/ferent quantum gravitational models.

\section{Discussion}
\label{Dis}
As a simplest radiation object, the black holes can be considered as a burning coal, which thermal radiation can be described perfectly well by Planck's law.  However, many works claimed that the property of  thermal radiation of microscopic objects is quite dif\/ferent from that of macroscopic objects \cite{chd1,chd2,chd3}. Therefore, it is necessary to f\/ind a modif\/ied theory to describe behavior of thermal radiation of soot particles or a black hole that approach the Planck scale. In the present work, we have investigated the quantum gravity corrections to information f\/lux of SC black hole and its sparsity via the rainbow functions that have been proposed by Amelino-Camelia, \emph{et al}. First, by using the thermodynamics quantities of rainbow SC black hole, we found a new relationship between the mass and Bekenstein entropy loss per emitted quanta, which implies that the information f\/lux of rainbow SC black hole varies with its mass. When rainbow SC black hole approaches the Planck scale, the information f\/lux would reduce to zero. Accordingly, the ef\/fect of RG can stop the evaporation of black hole and leads to a remnant. Hence, one can study the lifetime of rainbow SC black hole via its information f\/lux. Moreover, sparsity of Hawking radiation has also been analyzed. The results showed that the sparsity of Hawking radiation is no longer a constant; instead, it monotonically decreases as the mass of black hole decrease. From f\/igure~\ref{fig3}, one can see that the modif\/ied sparsity diverges as  $M \to M_{{\rm{res}}}$, which indicates that the pause time between in Hawking radiation becomes longer and longer. Finally, it is also found that the modif\/ied dimensionless parameter $\chi$  in our work is  remains qualitatively the same as the results in ref. \cite{chc2}, which contains a negative GUP parameter. However, when the GUP parameter takes a positive value, the GUP corrected $\chi$ is dif\/ferent from our case. Those dif\/ferences may be caused by the dif\/ferent models of quantum gravity. Actually, there are a lot of works try to investigate the relationship between GR and GUP since they are able to inf\/luence the evaporation process of a black hole, and our work just showed the similarities and dif\/ferences between the RG and GUP from the perspective of information loss.

\section*{Acknowledgements}
This work is  supported in part by the National Natural Science Foundation of China (Grant Nos. 11847048 and 11573022) and the Fundamental Research Funds of China West Normal University (Grant Nos. 17E093 and 17YC518).


\begin{thebibliography}{99}
\bibitem{cha1}
F. R. Tangherlini, ILNuovo Cimento \textbf{27}, (1963) 636. \href{https://doi.org/10.1007/BF02784569} {DOI: 10.1007/BF02784569}

\bibitem{cha2}
J. M. Bardeen, B. Carter, S. W. Hawking,  Commun.\ Math.\ Phys.\ \textbf{31}, (1973) 161. \href{https://doi.org/10.1007/BF01645742} {DOI: 10.1007/BF01645742}

\bibitem{cha3}
J. D. Bekenstein, Lett.\ Nuovo.\ Cim.\ \textbf{4}, (1972) 737. \href{https://doi.org/10.1007/BF02757029} {DOI: 10.1007/BF02757029}

\bibitem{cha4}
J. D. Bekenstein, Phys.\ Rev.\ D \textbf{7}, 2333 (1973). \href{https://doi.org/10.1103/PhysRevD.7.2333} {DOI: 10.1103/PhysRevD.7.2333}

\bibitem{cha5}
S. W. Hawking, Commun.\ Math.\ Phys.\ \textbf{43}, 199 (1975). \href{https://doi.org/10.1007/BF02345020} {DOI: 10.1007/BF02345020}

\bibitem{cha6}
S. W. Hawking, Nature, \textbf{248}, (1974) 30. \href{https://doi.org/10.1038/248030a0} {DOI: 10.1038/248030a0}

\bibitem{cha7}
K. Srinivasan, T. Padmanabhan, Phys.\ Rev.\ D \textbf{60}, (1999) 24007. \href{https://arxiv.org/abs/gr-qc/9812028} {arXiv:gr-qc/9812028}

\bibitem{cha8}
T. Damour, R. Runi, Phys.\ Rev.\ D \textbf{14}, (1976) 332. \href{https://doi.org/10.1103/PhysRevD.14.332} {DOI: 10.1103/PhysRevD.14.332}

\bibitem{cha9}
M. K. Parikh, F. Wilczek, Phys.\ Rev.\ Lett.\ \textbf{85}, (2000) 5042. \href{https://arxiv.org/abs/hep-th/9907001} {arXiv:hep-th/9907001}

\bibitem{cha10}
R. Banerjee, B. R. Majhi, Phys.\ Lett.\ B \textbf{662}, (2008) 62. \href{https://arxiv.org/abs/0801.0200} {arXiv:0801.0200}

\bibitem{cha11}
J. D. Bekenstein, Phys.\ Rev.\ D \textbf{51}, (1995) 6608. \href{https://doi.org/10.1103/PhysRevD.51.R660} {DOI: 10.1103/PhysRevD.51.R660}

\bibitem{cha12}
D. N. Page, Phys.\ Rev.\ D \textbf{13}, (1976) 198. \href{https://doi.org/10.1103/PhysRevD.13.198} {DOI: 10.1103/PhysRevD.13.198}

\bibitem{cha13}
D. N. Page, Phys.\ Rev.\ D \textbf{14}, (1976) 3260. \href{https://doi.org/10.1103/PhysRevD.14.3260} {DOI: 10.1103/PhysRevD.14.3260}

\bibitem{cha14}
D. N. Page, Phys.\ Rev.\ D \textbf{16}, (1977) 2402. \href{https://doi.org/10.1103/PhysRevD.16.2402} {DOI: 10.1103/PhysRevD.16.2402}

\bibitem{cha15}
A. Alonso-Serrano, M. Visser, Entropy \textbf{19}, (2017) 207. \href{https://arxiv.org/abs/1704.00237} {arXiv:1704.00237}

\bibitem{cha16}
A. Alonso-Serrano, M. Visser, Phys. Lett. B \textbf{757}, (2016) 383. \href{https://arxiv.org/abs/1511.01162} {arXiv:1511.01162}

\bibitem{cha17}
A. Alonso-Serrano, M. Visser, Phys.\ Lett.\ B \textbf{776}, (2018) 10. \href{https://doi.org/10.1016/j.physletb.2017.11.020} {DOI: 10.1016/j.physletb.2017.11.020}

\bibitem{cha18}
A. Almheiri, D. Marolf, J. Polchinski, J. Sully, JHEP \textbf{02}, (2013) 062. \href{https://arxiv.org/abs/1207.3123} {arXiv:1207.3123}

\bibitem{cha19}
S. Carlip, Int.\ J.\ Mod.\ Phys.\ D \textbf{23}, (2014) 1430023. \href{https://arxiv.org/abs/1410.1486} {arXiv:1410.1486}

\bibitem{cha20}
P. Chen, Y. C. Ong, D.-H. Yeom, Phys.\ Rep.\ \textbf{603}, (2015) 1. \href{https://arxiv.org/abs/1412.8366} {arXiv:1412.8366}

\bibitem{cha21}
D. Marolf, Rep. Prog. Phys. \textbf{80}, (2017) 092001. \href{https://arxiv.org/abs/1703.02143} {arXiv:1703.02143}

\bibitem{cha22}
A. Almheiri, D. Marolf, J. Polchinski, D. Stanford, J. Sully, JHEP \textbf{09}, (2013) 018. \href{https://arxiv.org/abs/1304.6483} {arXiv:1304.6483}

\bibitem{cha23}
J. Maldacena, L. Susskind, Fortsch.\ Phys.\ \textbf{61}, 781 (2013). \href{https://arxiv.org/abs/1306.0533} {arXiv:1306.0533}

\bibitem{cha24}
P. Chen, Y. C. Ong, D. N. Page, M. Sasaki, D. H. Yeom, Phys. Rev. Lett. \textbf{116}, (2016) 161304. \href{https://arxiv.org/abs/1511.05695} {arXiv:1511.05695}

\bibitem{cha25}
W. G. Unruh, R. M. Wald, Rep. Prog. Phys. \textbf{80}, (2017) 092002. \href{https://arxiv.org/abs/1703.02140} {arXiv:1703.02140}

\bibitem{cha26}
R. J. Adler, P. Chen, D. I. Santiago, Gen. Relativ. Gravit. \textbf{33}, (2001) 2101. \href{https://arxiv.org/abs/gr-qc/0106080} {arXiv:gr-qc/0106080}

\bibitem{cha27+}
C. D. Chen, H. W. Wu, H. T. Yang, S. Z Yang, Int. J Mod. Phys. A \textbf{29}, (2014) 1430054. \href{https://doi.org/10.1142/S0217751X14300543} {DOI: 10.1142/S0217751X14300543}

\bibitem{cha27}
Z. W. Feng, H. L. Li, X. T. Zu, S. Z. Yang, Eur. Phys. J. C \textbf{76}, (2016) 212. \href{https://arxiv.org/abs/1604.04702} {arXiv:1604.04702}

\bibitem{cha28}
X.-Q. Li, Phys.\ Lett.\ B \textbf{763}, (2016) 80. \href{https://arxiv.org/abs/1605.03248} {arXiv:1605.03248}

\bibitem{cha29}
H.-L. Li, S.-R. Chen, Gen. Rel. Grav. \textbf{49}, (2017) 128. \href{https://doi.org/10.1007/s10714-017-2296-6} {DOI: 10.1007/s10714-017-2296-6}

\bibitem{cha30}
S. Gangopadhyay, A. Dutta, Adv. High Energy Phys. \textbf{2018}, (2018) 7450607. \href{https://arxiv.org/abs/1805.11962} {arXiv:1805.11962}

\bibitem{cha31}
A. Alonso-Serrano, M. P. D\c{a}abrowski, H. Gohar, Phys.\ Rev.\ D \textbf{97}, (2018) 044029. \href{https://arxiv.org/abs/1801.09660} {arXiv:1801.09660}

\bibitem{cha31+}
A. Alonso-Serrano, M. P. D\c{a}abrowski, H. Gohar, Int. J. Mod. Phys. D \textbf{27}, (2018)  847028. \href{https://doi.org/10.1142/S0218271818470284} {DOI: 10.1142/S0218271818470284}

\bibitem{cha32}
G. Amelino-Camelia, Int. J. Mod. Phys. D \textbf{11}, (2002)  35. \href{https://arxiv.org/abs/gr-qc/0012051} {arXiv:gr-qc/0012051}

\bibitem{cha33}
J. Magueijo, L. Smolin,  Class. Quant. Grav. \textbf{21}, (2004) 1725. \href{https://arxiv.org/abs/gr-qc/0305055} {arXiv:gr-qc/0305055}

\bibitem{cha34}
A. F. Ali, Phys.\ Rev.\ D \textbf{89}, (2014) 104040. \href{https://arxiv.org/abs/1402.5320} {arXiv:1402.5320}

\bibitem{cha34a1}
A. F. Ali, M. Faizal, M. M. Khalil, Nucl.Phys. B \textbf{894}, (2015) 341. \href{https://arxiv.org/abs/1410.5706} {arXiv:1410.5706}

\bibitem{cha34a2}
A. F. Ali, M. Faizal, M. M. Khalil, JHEP \textbf{1412}, (2014) 159. \href{https://arxiv.org/abs/1409.5745} {arXiv:1409.5745}

\bibitem{cha34a3}
A. F. Ali, M. Faizal, M. M. Khalil, Phys.Lett. B  \textbf{743}, (2015) 295. \href{https://arxiv.org/abs/1410.4765} {arXiv:1410.4765}

\bibitem{cha34+}
C. Rovelli, F. Vidotto, Universe, \textbf{4}, (2018), 127.  \href{https://arxiv.org/abs/1805.03872} {arXiv:1805.03872}

\bibitem{cha35+}
C. Rovelli, F. Vidotto, Int. J Mod. Phys. D \textbf{23}, (2014) 1442026. \href{https://arxiv.org/abs/1401.6562} {arXiv:1401.6562}

\bibitem{cha36}
Y. Gim, W. Kim, JCAP \textbf{05}, (2015) 002. \href{https://arxiv.org/abs/1501.04702} {arXiv:1501.04702}

\bibitem{cha37}
S. H. Hendi, M. Faizal, Phys.\ Rev.\ D \textbf{92}, (2015) 044027. \href{https://arxiv.org/abs/1506.08062} {arXiv:1506.08062}

\bibitem{cha38}
S. H. Hendi, M. Faizal, B. E. Panah, S. Panahiyan, Eur. Phys. J. C \textbf{76}, (2016) 296. \href{https://arxiv.org/abs/1508.00234} {arXiv:1508.00234}

\bibitem{cha39}
S. H. Hendi, B. Eslam Panah, S. Panahiyan, Phys.\ Lett.\ B \textbf{769}, (2017) 191. \href{https://arxiv.org/abs/1602.01832} {arXiv:1602.01832}

\bibitem{cha40}
S. H. Hendi, S. Panahiyan, B. E. Panah,  M. Momennia, Eur. Phys. J. C \textbf{76}, (2016) 150. \href{https://arxiv.org/abs/1512.05192} {arXiv:1512.05192}

\bibitem{cha40a1}
S. H. Hendi, S. Panahiyan, B. E. Panah,  Adv. High Energy Phys. \textbf{2015},  (2015) 743086. \href{https://arxiv.org/abs/1509.07014 } {arXiv:1509.07014 }

\bibitem{cha40a2}
S. H. Hendi, G. H. Bordbar, B. Eslam Panah, S. Panahiyan, JCAP \textbf{09},  (2016) 013. \href{https://arxiv.org/abs/1509.05145} {arXiv:1509.05145}

\bibitem{cha40a3}
S. H. Hendi, G.-H. Li, J.-X. Mo, S. Panahiyan, B. E. Panah,  Eur. Phys. J. C \textbf{76},  (2016) 571. \href{https://arxiv.org/abs/1608.03148} {arXiv:1608.03148}

\bibitem{cha40a5}
S. H. Hendi, S. Panahiyan, S. Upadhyay, B. Eslam Panah,  Phys. Rev. D \textbf{95},  (2017) 084036.  \href{https://arxiv.org/abs/1611.02937} {arXiv:1611.02937}

\bibitem{cha40a6}
S. H. Hendi, B. Eslam Panah, S. Panahiyan, M. Momenna,   Eur. Phys. J. C \textbf{77},  (2017) 647. \href{https://arxiv.org/abs/1708.06634} {arXiv:1708.06634}

\bibitem{cha41}
Y. Gim, W. Kim, Eur. Phys. J. C \textbf{76}, (2016) 166. \href{https://arxiv.org/abs/1509.06846} {arXiv:1509.06846}

\bibitem{cha42}
H.-L. Li, Z.-W. Feng, S.-Z. Yang, X.-T. Zu, Eur. Phys. J. C \textbf{78} (2018) 768. \href{https://doi.org/10.1140/epjc/s10052-018-6252-8} {DOI:10.1140/epjc/s10052-018-6252-8}

\bibitem{cha44}
B. Eslam Panah, Phys.\ Lett.\ B \textbf{787}, (2018) 45.  \href{https://arxiv.org/abs/1805.03014} {arXiv:1805.03014}

\bibitem{cha40a4}
S. H. Hendi, M. Momennia, B. E. Panah, M. Faizal,  Astrophys. J \textbf{827},  (2016) 153.  \href{https://arxiv.org/abs/1703.00480} {arXiv:1703.00480}

\bibitem{cha40a7}
S. H. Hendi, M. Momennia, B. E. Panah, S. Panahiyan, Phys. Dark Universe \textbf{16},  (2017) 26. \href{https://arxiv.org/abs/1705.01099} {arXiv:1705.01099}

\bibitem{cha40b1}
S. H. Hendi, G. H. Bordbar, B. Eslam Panah, S. Panahiyan,  JCAP \textbf{07}, (2017) 004. \href{https://arxiv.org/abs/1701.01039} {arXiv:1701.01039}

\bibitem{cha40b2}
S. H. Hendi, G. H. Bordbar, B. Eslam Panah, S. Panahiyan, JCAP \textbf{09}, (2016) 013. \href{https://arxiv.org/abs/1509.05145} {arXiv:1509.05145}

\bibitem{cha40b5}
Cl\'{e}sio E. Mota, Luis C. N. Santos, Guilherme Grams, Franciele M. da Silva, D\'{e}bora P. Menezes, Phys. Rev. D \textbf{100}, 024043 (2019). \href{https://arxiv.org/abs/1905.01250} {arXiv:1905.01250}

\bibitem{cha40b3}
Y. Gim, B. Gwak, Phys. Lett. B \textbf{794}, (2019) 122.  \href{https://arxiv.org/abs/1808.05943 } {arXiv:1808.05943 }

\bibitem{cha40b4}
P. Channuie, Eur. Phys. J. C \textbf{79}, (2019) 508.  \href{https://arxiv.org/abs/1903.05996} {arXiv:1903.05996}

\bibitem{chb0}
G. Amelino-Camelia, J. R. Ellis, N. Mavromatos, D. V. Nanopoulos, S. Sarkar, Nature \textbf{393}, (1998) 763. \href{https://arxiv.org/abs/astro-ph/9712103} {arXiv:astro-ph/9712103}

\bibitem{chb0+}
J. Magueijo, L. Smolin, Phys.\ Rev.\ Lett.\ \textbf{88}, (2002) 190403. \href{https://arxiv.org/abs/hep-th/0112090} {arXiv:hep-th/0112090}

\bibitem{chb2+}
A. F. Ali, M. M. Khalil, Europhys.\ Lett.\ \textbf{110}, (2015) 20009. \href{https://arxiv.org/abs/1408.5843} {arXiv:1408.5843}

\bibitem{chb1}
G. Amelino-Camelia, Living Rev.\ Rel.\ \textbf{16}, (2013) 5. \href{https://arxiv.org/abs/0806.0339} {arXiv:0806.0339}

\bibitem{chb2}
G. Amelino-Camelia, J. R. Ellis, N. E. Mavromatos, D. V. Nanopoulos, Int.\ J.\ Mod.\ Phys.\ A \textbf{12}, (1997) 607. \href{https://arxiv.org/abs/hep-th/9605211} {arXiv:hep-th/9605211}

\bibitem{chb2a+}
B. Gwak, W. Kim, B.-H. Lee, \href{https://arxiv.org/abs/1608.04247} {arXiv:1608.04247}

\bibitem{chb3}
A. F. Ali, M. Faizal, M. M. Khalil, JHEP \textbf{1412}, (2014) 159. \href{https://arxiv.org/abs/1409.5745} {arXiv:1409.5745}

\bibitem{chb4}
M. Shahjalal, Phys.\ Lett.\ B  \textbf{784}, (2018) 6. \href{https://doi.org/10.1016/j.physletb.2018.07.032} {DOI: 10.1016/j.physletb.2018.07.032}

\bibitem{chb4+}
G. Yadav, B. Komal, B. R. Majhi, Int.\ J Mod.\ Phys.\ A \textbf{32}, (2017) 1750196. \href{https://arxiv.org/abs/1605.01499} {arXiv:1605.01499}

\bibitem{chb5+}
A. Paul, B. R. Majhi,  Int.\ J.\ Mod.\ Phys.\ A \textbf{32}, (2017) 1750088. \href{https://arxiv.org/abs/1601.07310} {arXiv:1601.07310}

\bibitem{chb6+}
G. Yadav, B. Komal, B. R. Majhi.   Int. J. Mod. Phys. A \textbf{32}, (2017)1750196.   \href{https://arxiv.org/abs/1605.01499} {arXiv:1605.01499}

\bibitem{chb7+}
Z.-W. Feng, Q.-C. Ding, S.-Z. Yang, Eur. Phys. J. C \textbf{79}, (2019) 45 .   \href{https://arxiv.org/abs/1810.05645} {arXiv:1810.05645}

\bibitem{chb5}
G. Amelino-Camelia, M. Arzano, A. Procaccini, Phys.\ Rev.\ D \textbf{70}, (2004) 107501. \href{https://arxiv.org/abs/gr-qc/0405084} {arXiv:gr-qc/0405084}

\bibitem{chb6}
B. Mu, P. Wang, H. Yang, JCAP. \textbf{11}, (2015) 045. \href{https://arxiv.org/abs/1507.03768} {arXiv:1507.03768}

\bibitem{chc1}
F. Gray, S. Schuster, A. Van-Brunt, M. Visser, Class.\ Quant.\ Grav.\ \textbf{33}, (2016) 115003. \href{https://arxiv.org/abs/1506.03975} {arXiv:1506.03975}

\bibitem{chc2}
Y. C. Ong, JHEP \textbf{10}, (2018) 195. \href{https://arxiv.org/abs/1806.03691} {arXiv:1806.03691}

\bibitem{chc3}
Y. C. Ong, JCAP \textbf{9}, (2018) 15. \href{https://arxiv.org/abs/1804.05176} {arXiv:1804.05176}

\bibitem{chc4}
M. Moussa, Adv.  High Energy Phys.  \textbf{2015}, (2018) 343284. \href{https://arxiv.org/abs/1512.04337} {arXiv:1512.04337}

\bibitem{chc5}
 R. Garattini, G. Mandanici, Eur. Phys. J. C \textbf{77}, (2017)  57.  \href{https://arxiv.org/abs/1607.08234} {arXiv:1607.08234}

\bibitem{chc6}
F. Lu, J. Tao, P. Wang, JCAP \textbf{2018}, (2018) 36. \href{https://arxiv.org/abs/1811.02140} {arXiv:1811.02140}

\bibitem{chd1}
X. Liu, T. Tyler, T. Starr, A. F. Starr, N. M. Jokerst, W. J. Padilla, Phys. Rev. Lett. \textbf{107}, (2011) 045901. \href{https://doi.org/	10.1103/PhysRevLett.107.045901} {DOI: 10.1103/PhysRevLett.107.045901}

\bibitem{chd2}
C. Wuttke, A. Rauschenbeutel, Phys. Rev. Lett. \textbf{111}, (2013) 024301.  \href{https://arxiv.org/abs/1209.0536} {arXiv:1209.0536}

\bibitem{chd3}
D. Thompson, L. Zhu, R. Mittapally, S. Sadat, Z. Xing, P. McArdle, M. M. Qazilbash, P. Reddy, E. Meyhofer, Nature \textbf{561},  (2018) 216. \href{https://doi.org/10.1038/s41586-018-0480-9} {DOI: 10.1038/s41586-018-0480-9}

\end{thebibliography}
\end{document}